\documentclass[aps,prb,twocolumn,superscriptaddress]{revtex4}
\usepackage{epsfig,pslatex,latexsym,times,amssymb,amsmath,graphicx}
\usepackage{bm}
\usepackage{amsmath}
\usepackage{amsfonts}
\usepackage{amssymb}
\usepackage{graphicx}
\providecommand{\U}[1]{\protect\rule{.1in}{.1in}}
\begin{document}
\author{N.J. Harmon}
\affiliation{Physics Department, Ohio State University, 191 W. Woodruff Ave., Columbus, OH
43210, USA}
\author{W.O. Putikka}
\affiliation{Physics Department, Ohio State University, 191 W. Woodruff Ave., Columbus, OH
43210, USA}
\author{R. Joynt}
\affiliation{Physics Department, University of Wisconsin-Madison, 1150 University Ave., WI
53705, USA}
\date{\today}
\title{Theory of Electron Spin Relaxation in ZnO}

\begin{abstract}
Doped ZnO is a promising material for spintronics applications. \ For such
applications, it is important to understand the spin dynamics and particularly
the spin relaxation times of this II-VI semiconductor. \ The spin relaxation
time $\tau_{s}$ has been measured by optical orientation experiments, and it
shows a surprising non-monotonic behavior with temperature. \ We explain this
behavior by invoking spin exchange between localized and extended states.
\ Interestingly, the effects of spin-orbit \ coupling are by no means
negligible, in spite of the relatively small valence band splitting. \ This is
due to the wurtzite crystal structure of ZnO. \ Detailed analysis allows us to
characterize the impurity binding energies and densities, showing that optical orientation experiments can be used as a
characterization tool for semiconductor samples.

\end{abstract}
\maketitle

\section{Introduction}

Zinc Oxide has been the subject of considerable experimental and theoretical
investigation for many years.\cite{ozgur} \ Its bandgap is in the near
ultraviolet, making it useful as a transparent conductor and as sunscreen.
\ Its piezoelectricity opens up transduction applications. \ The activity has
intensified more recently because of the possibility that ZnO might be useful
for spintronics or spin-based quantum computation. \ It has been predicted to
be a room-temperature ferromagnet when doped with Mn.\cite{dietl}
\ Furthermore, its spin-orbit coupling is generally thought to be very weak
compared with GaAs. \ The usual measure of the strength of spin-orbit coupling
in semiconductors is the energy splitting at the top of the valence band. \ It
is said that the spin-orbit coupling is negligible in ZnO because the
valence-band splitting is $-3.5$ meV,\cite{fan} as opposed to $340$ meV for
GaAs. \ Smaller spin-orbit coupling should lead to long spin relaxation times.
\ Long relaxation times are required if spin information is to be transported
over appreciable distances.

The spin relaxation time $\tau_{s}$ has been measured by Ghosh et al.
\cite{ghosh} to be about 20 ns from 0 to 20 K in optical orientation
experiments. $\tau_{s}$ is sometimes called $T_{2}^{\ast}$ even in the absence
of an external field. Since the data from Ref. (\onlinecite{ghosh}) used in
this paper were taken at zero field, the relaxation time is taken to be
$\tau_{s}$ to avoid confusion with experiments conducted at finite field.
\ The data show two surprising features. \ First, the relaxation times are
actually somewhat shorter than the longest relaxation times in GaAs, which are
about 100 ns.\cite{kikkawa} \ One might expect the opposite given the relative
strength of spin-orbit coupling in the two materials. \ \ Second, $\tau_{s}$
shows a non-monotonic temperature dependence, first increasing slightly and
then rapidly decreasing - but increasing temperature usually promotes spin relaxation.

In this paper, we show that a theory previously developed for $\tau_{s}$ in
GaAs \cite{putikka} can account for these observations. \ The theory must be
modified to take account of the different impurity levels and binding energies
of ZnO. \ This is important, because, in spite of intensive investigation, the
nature of the impurities that govern the electrical properties of ZnO remains
controversial, and our analysis sheds some light on this issue. \ Even more
interestingly, it turns out that the wurtzite crystal structure has very
important consequences for the D'yakonov-Perel (DP) \cite{dyakonov} scattering
that dominates the relaxation at higher temperatures. \ Thus the crystal
structure must be taken into account fully. \ The final message will be that
the \textquotedblleft weak\textquotedblright\ spin-orbit coupling of ZnO is
not negligible for spin relaxation, and it does not lead to long relaxation
times. \ 

In the next section we give the background information for ZnO. \ Sec. III is
devoted to a derivation of the equations of motion for the spins. \ In Sec. IV
the computational method is described. \ The determination of the parameters
in the equations of motion is a separate task. \ The most important of the
parameters is that which controls that DP spin relaxation. \ Since the
calculation of this parameters is not straightforward, we devote Sec. V to
that. \ Sec. VI gives the results. \ Sec. VII is the conclusions and puts the
results into context.

\section{Background}

\ In ZnO produced by the hydrothermal method, it is generally thought that
there are two sets of impurity states, one shallow and quasi-hydrogenic, one
deep and very well localized.\cite{wenckstern, grossner} \ Their precise
physical nature is not known. \ In the case of the deep impurity, it is
believed that a lattice defect accompanies the chemical impurity. \ The
binding energies are in the range of a few 10s of meV for the shallow impurity
and a few 100s of meV for the deep impurity. \ We shall demonstrate below that
the optical orientation data can put bounds on these numbers.

ZnO crystallizes in the wurtzite structure rather than the zincblende
structure familiar from the III-V compounds. \ This has very important
implications for the conduction band states. \ The spin-orbit interaction
lifts the spin degeneracy in the conduction band. \ In zincblende structures
crystal symmetry implies that the splitting is cubic in the magnitude of the
wavevector $k$, but in the wurtzite structure the splitting is
linear.\cite{voon} \ However, the spin relaxation time of the low-lying
conduction band states depends mainly on the spin splitting near the
conduction band minimum, and this is larger in ZnO than in GaAs for small
enough $k$.

In optical orientation experiments, electrons are excited from the valence
band to the conduction band by circularly polarized light tuned close to the
bandgap energy (pump step). \ The population of conduction electrons so
created is spin-polarized.\cite{oo2} \ Energy relaxation then occurs on a
short time scale ($\leq1$ ns), but most of this relaxation is from
spin-conserving processes, so there is a longer time scale (or time scales) on
which the spin of the system relaxes. \ When an external magnetic field is
present, the time scale to relax the transverse component of the
\textit{net} magnetization is called $\tau_{s}$. \ It is measured using
Faraday rotation or the Kerr effect (probe step).

The important physical point is that the fast energy relaxation leads to a
thermal charge distribution for the electrons by the time 1 ns has elapsed,
but the spin distribution relaxes on longer time scales. \ The thermal charge
distribution means that the localized donor impurity states are mostly full at
the relatively low temperatures of the experiment. \ \ The spins of the
localized electrons must be included along with the conduction electron spins.
\ The spins of localized and extended states can be interchanged by the
exchange coupling, a process we call cross- relaxation. \ This is often a
rather fast process and is particularly important when the relaxation times of
the localized and extended states are very different in magnitude. \ In GaAs
this process is important in all the regimes of temperature, applied field,
and impurity density that have been studied, and it is important in ZnO as well.

In the following section we derive a set of modified Bloch equations to
describe the aforementioned spin dynamics.

\section{Modified Bloch Equations\ \ }

We consider a conduction electron in the semiclassical approximation. \ It
moves as a wavepacket with a well-defined momentum and scatters from
impurities and phonons at time intervals of average length $\tau_{p},$ where
$\tau_{p}$ is the momentum relaxation time. \ \ Its spin operator is
$\mathbf{s}_{c}.$ The spin-dependent part of its Hamiltonian in the absence of
an external magnetic field is:
\begin{equation}
H^{c}=H_{1}^{c}+H_{2}^{c}=-\frac{1}{2\hslash}\sum\limits_{i}J\left(
\mathbf{r}-\mathbf{R}_{i}\right)  \mathbf{s}_{i}\cdot\mathbf{s}_{c}-g\frac
{\mu_{B}}{\hbar}\mathbf{b}\left(  t\right)  \cdot\mathbf{s}_{c}.
\end{equation}
The first term, $H_{1}^{c}$, is the exchange interaction with impurity spins
$\mathbf{s}_{i}$ located at an positions $\mathbf{R}_{i}$. \ It is the same
interaction that is responsible for the Kondo effect, but the temperatures
here are all much greater than the Kondo temperature. \ The range of the
function $J\left(  \mathbf{r}-\mathbf{R}_{i}\right)  $ is roughly $a_{B},$
where $a_{B}$ is the effective Bohr radius. \ The second term, $H_{2}^{c}$,
represents other spin relaxation mechanisms that we model as a small random
classical field $\mathbf{b}(t)$ with a correlation time much shorter than
$\tau_{s}$. An analogous Hamiltonian $H^{l}$ can be written for a localized electron.

First, we concentrate on the spin dynamics resulting from the spin-spin term
and ignore the second term. In the dilute limit ($a_{B}n_{imp}^{1/3}\ll1$), a
conduction electron encounters impurities with randomly aligned spins if no
short-range order is present in the impurity system. An effective field from
the impurity spin affects the conduction electron when it is within $~\sim
a_{B}$ of the impurity. When $|\mathbf{r}-\mathbf{R}_{i}|>a_{B}$, the
conduction electron proceeds unhindered by the effective field. This effective
field is a result of the exchange potential. An itinerant electron will spend
an average time of $a_{B}/v$ within the range of the effective field where $v$
is the velocity of the electron. Thus the time between encounters\cite{seeger}
is $1/n_{l}a_{B}^{2}v$.

In a semiclassical picture the spin of the itinerant electron undergoes
precession of magnitude $\Delta\phi= J a_{B}/2 v$ through a random angle
during each encounter with an impurity. The spin of the impurity electron also
precesses but with angle $-\Delta\phi$. Since the sum of spins, $\mathbf{s}
_{c} + \mathbf{s}_{l}$, commutes with $H_{1}^{c} + H_{1}^{l}$, the total spin
in the system must be conserved. However the spin in each subsystem may shift
between one another; this is cross-relaxation.

It turns out for the parameters of the system under consideration that
$\Delta\phi\sim1$, and we then find that
\begin{equation}
\tau_{c}^{cr}\sim\frac{1}{n_{l}a_{B}^{2}v}%
\end{equation}
which implies that the spin is essentially randomized after one impurity encounter.

If we consider an ensemble of conduction electrons with a net magnetization
$m_{c}$, this magnetization is exchanged at a rate of $1/\tau_{c}^{cr}$. As
previously mentioned, any magnetization lost from the conduction electrons
must be gained by the localized electrons and vice-versa. For clarity we write
$1/\tau_{c}^{cr}=n_{l}/\gamma^{cr}$ and $1/\tau_{l}^{cr}=n_{c}/\gamma^{cr}$
where $\gamma^{cr}=1 / a_{B}^{2}v$.

We now examine the second term of the Hamiltonian
\begin{equation}
H_{2}^{c}\left(  t\right)  =-\frac{1}{2}g\mu_{B}\big(b_{x}{\small (t)\sigma
_{x}+b_{y}\left(  t\right)  \sigma_{y}+b_{z}\left(  t\right)  \sigma
_{z}\big).}%
\end{equation}

This Hamiltonian relaxes the conduction electron spin. To extract a relaxation
rate from this Hamiltonian, we use the equation of motion
\begin{equation}
\frac{d\rho(t)}{dt}=\frac{i}{\hbar}[\rho(t),H_{2}^{c}(t)]
\end{equation}
where $\rho(t)$ is the $2\times2$ spin density matrix for an electron of a
given momentum. \ We assume that the total density matrix for the conduction
electron factorizes; we neglect off-diagonal terms that come from
correlations.  By iteration, we can write this equation as
\begin{align}
&  \left\langle \frac{d\rho\left(  t\right)  }{dt}\right\rangle
=\nonumber\label{vonNeumann}\\
&  \frac{i}{\hbar}\left\langle [\rho\left(  0\right)  ,H_{2}^{c}\left(
t^{\prime}\right)  ]\right\rangle -\frac{1}{\hbar^{2}}\int_{0}^{t}\left\langle
{\left[  \left[  \rho\left(  t^{\prime}\right)  ,H_{2}^{c}\left(  t^{\prime
}\right)  \right]  ,H_{2}^{c}\left(  t\right)  \right]  }\right\rangle
dt^{\prime}%
\end{align}
where the angular brackets indicate averaging over all orientations of $b(t)$.
To simplify notation, from now on angular brackets will be suppressed on the
density matrix. Since $\langle b_{i}(t)\rangle=0$, the first term is zero. We
assume that  different directions of $b_{i}$ are uncorrelated and (since the
external field is zero) different direction are equivalent. \ Then we have
$\langle b_{i}(t)b_{j}(t^{\prime})\rangle=\langle b(t)b(t^{\prime}
)\rangle\delta_{i,j}.$  Therefore, Eq. (\ref{vonNeumann}) reduces to
\begin{equation}
\frac{d\rho\left(  t\right)  }{dt}=-\frac{g^{2}\mu_{B}^{2}}{2\hbar^{2}}%
\int_{0}^{t}\sum_{i}[\rho(t^{\prime}),\sigma_{i}]\sigma_{i}\langle
b(t)b(t^{\prime})\rangle dt^{\prime}.\label{vonNeumann2}%
\end{equation}
The correlation function is assumed to be stationary in time so $\langle
b(t)b(t^{\prime})\rangle=g(t^{\prime}-t)=g(\tau)$.\cite{slichter} If the
correlation time of the $b$-fluctuations, $\tau_{e}$, is short, $\rho$ will
not change on that timescale and $g(\tau)$ will be nearly a $\delta$- function
(Markov approximation). Eq. (\ref{vonNeumann2}) can then be written as
\begin{equation}
\frac{d\rho\left(  t\right)  }{dt}=-\frac{2g^{2}\mu_{B}^{2}}{\hbar^{2}}%
\frac{1}{4}\sum_{i}[\rho(t),\sigma_{i}]\sigma_{i}\int_{0}^{\infty}\left\langle
b(t)b(t^{\prime})\right\rangle dt^{\prime}.
\end{equation}
The integral is approximated by $\langle b^{2}\rangle\tau_{e}$. Define the
relaxation time scale $\tau_{c}$ by
\begin{equation}
\frac{1}{\tau_{c}}=2\left(  \frac{g\mu_{B}}{\hbar}\right)  ^{2}\langle b^{2}\rangle\tau_{e}%
\end{equation}
giving
\begin{equation}
\frac{d\rho\left(  t\right)  }{dt}=-\frac{1}{4\tau_{c}}\sum_{i}[\rho
(t),\sigma_{i}]\sigma_{i}.\label{tc}%
\end{equation}
The density matrix can be expanded in Pauli spin matrices
\begin{equation}
\rho\left(  t\right)  =\frac{1}{2}I+\frac{1}{2}\sum\limits_{i}m_{i}\left(
t\right)  \sigma_{i}.\label{pauli}%
\end{equation}
where $I$ is the $2\times2$ identity matrix and $m_{i}=Tr(\sigma_{i}\rho)$ is
the expected value of the magnetization. Inserting Eq. (\ref{pauli}) in Eq.
(\ref{tc}) and matching coefficients of Pauli matrices gives a set of
equations for the dynamics of $\mathbf{m}$. For instance for conduction
electron magnetization $m_{c}$ in the $x$-direction, $dm_{c}/dt=Tr(\sigma
_{x}d\rho/dt)=-m_{c}/\tau_{c}$. As with $H_{1}^{c}$, similar expressions for
the localized magnetization $m_{l}$ can be found: $dm_{l}/dt=Tr(\sigma
_{x}d\rho/dt)=-m_{l}/\tau_{l}$.

By combining the effects of $H_{1} = H_{1}^{c}+H_{1}^{l}$ and $H_{2} =
H_{2}^{c}+H_{2}^{l}$, the modified Bloch equations for the magnetizations can
be expressed as
%{\setlength\arraycolsep{2pt}%
\begin{align}
\label{bloch0}\frac{d m_{c}}{dt}  &  = -\Big( \frac{1}{\tau_{c}} + \frac
{n_{l}}{\gamma^{cr}} \Big) m_{c} + \frac{n_{c}}{\gamma^{cr}} m_{l}\nonumber\\
\frac{d m_{l}}{dt}  &  = \frac{n_{l}}{\gamma^{cr}} m_{c} -\Big( \frac{1}%
{\tau_{l}} + \frac{n_{c}}{\gamma^{cr}} \Big) m_{l}.
\end{align}
for two spin systems - itinerant and localized spins. $\tau_{c}$ and $\tau
_{l}$ in Eq. (\ref{bloch0}) are now described in terms of well known
relaxation mechanisms which will be discussed in the next section. This model
was successfully applied to GaAs.\cite{putikka} For ZnO, these Bloch equations
are easily extended to account for the multiple-type impurities present.

\section{Method \ \ }

We now seek to write equations like those of Eq. (\ref{bloch0}) with regard
given to the two types of impurities in ZnO - shallow and deep. As mentioned
above, we find that the cross-relaxation is important to understand the data.
\ These rates come from the Kondo-like $J\mathbf{s}_{l}\cdot\mathbf{s}_{c}$
interaction between an impurity spin $\mathbf{s}_{l}$ and a conduction band
spin $\mathbf{s}_{c}$. \ An expression for $J$ in terms of tight-binding
parameters can be derived using the Schrieffer-Wolf
transformation.\cite{schrieffer} \ One expects that the cross-relaxation
between conduction and shallow donor electrons to be much more rapid than the cross-relaxation
between conduction and deep donor electrons because of the greater binding energy of the deep
impurity and its larger on-site Coulomb energy. \ This is confirmed by the fit
to the data. \ In fact we find that terms involving cross-relaxation
between the deep donors and either the conduction band electrons or the
shallow donor electrons can be neglected. With these simplifications, for ZnO
Eq. (\ref{bloch0}) extends to
%{\setlength\arraycolsep{2pt}%
\begin{align}
\label{eq:blochSimp}\frac{d m_{c}}{dt}  &  = -\Big( \frac{1}{\tau_{c}} +
\frac{n_{ls}}{\gamma^{cr}_{c,s}} \Big) m_{c} + \frac{n_{c}}{\gamma^{cr}_{c,s}}
m_{ls}\nonumber\\
\frac{d m_{ls}}{dt}  &  = \frac{n_{ls}}{\gamma^{cr}_{c,s}} m_{c}
-\Big( \frac{1}{\tau_{ls}} + \frac{n_{c}}{\gamma^{cr}_{c,s}} \Big) m_{ls}\\
\frac{d m_{ld}}{dt}  &  = -\frac{1}{\tau_{ld}} m_{ld} .\nonumber
\end{align}
In this equation, $m_{c},$ $m_{ls},$ and $m_{ld}$ stand for the magnetizations
of the conduction electrons, the electrons on shallow impurities, and the
electrons on deep impurities, respectively. \ The $n$'s denote the
corresponding volume densities. \ \ Each of the populations has a relaxation
time $\tau_{c},\tau_{ls},$ and $\tau_{ld}.$
From Eq. (\ref{eq:blochSimp}), we can
then find the magnetization as a function of time.

Standard methods can be used to solve these differential equations. The
solutions yield a time dependence of the total magnetization, $m(t) = m_{c}(t)
+ m_{ls}(t) + m_{ld}(t)$, to be a sum of three exponentials, $\exp(-\Gamma_{+}
t)$, $\exp(-\Gamma_{-} t)$, and $\exp(-\Gamma_{d} t)$ where
\begin{equation}
\Gamma_{\pm} = \frac{1}{2} \Bigg(\frac{1}{\tau_{c}} + \frac{1}{\tau_{ls}} +
\frac{n_{c} + n_{ls}}{\gamma^{cr}_{c,s}} \pm S \Bigg), ~ \Gamma_{d} = \frac
{1}{\tau_{ld}}%
\end{equation}
with $S$ given by
\begin{equation}
S = \sqrt{\Bigg(\frac{1}{\tau_{ls}} - \frac{1}{\tau_{c}}+\frac{n_{c} - n_{ls}%
}{\gamma^{cr}_{c,s}}\Bigg)^{2} + \frac{4 n_{c} n_{ls}}{\gamma^{cr ~ 2}_{c,s}}%
}.
\end{equation}
No net moment can exist on the deep donor sites since no moment is excited
into the deep states on account of them being significantly below the
conduction band, and no net moment cross relaxes into these states. Therefore
$\Gamma_{d}$ can be ruled out as being the observed relaxation rate. In the
regime that $(n_{ls} + n_{c})/\gamma^{cr}_{c,s} \gg1/\tau_{c}, ~1/\tau_{ls}$,
the rate $\Gamma_{+}$ simplifies to $(n_{c} + n_{ls})/\gamma^{cr}_{c,s}$ and
is very rapid and the rate $\Gamma_{-}$ is slower,
\begin{equation}
\Gamma_{-}=\frac{n_{c}}{n_{c}+n_{ls}}\frac{1}{\tau_{c}}+\frac{n_{ls}}%
{n_{c}+n_{ls}}\frac{1}{\tau_{ls}}. \label{eq:eigenvalue}%
\end{equation}
We fit the data with this equation and associate it with $\tau_{s}$. \ We see
that the relaxation rate depends on two factors: the thermodynamic occupations
of the shallow donors (the deep donors are always nearly full in the
temperature range studied here) and form of the relaxation rates for the
conduction and localized shallow states.

The densities can be computed using standard formulas from equilibrium
statistical mechanics, since we deal only with time scales long compared to
the fast energy relaxation scale. \ As a function of temperature $T$, the
ratio $n_{c}/n_{ls}$ naturally increases rapidly as $T\rightarrow
|\varepsilon_{ls}|/k_{B},$ where $\varepsilon_{ls}$ is the binding energy of
the shallow impurity. \ $|\varepsilon_{ld}|$ is so large that these states are
always occupied at the experimental temperatures, which range from 5K to 80K.

$\tau_{c}$ is fairly complicated to calculate because there are several
mechanisms that can relax the conduction electron spins. The simplest such
mechanism is the Elliot-Yafet (EY) process\cite{elliott} that arises from spin
mixing in the wavefunctions. When a conduction electron is scattered by a
spin-independent potential from state $\mathbf{k}$ to state $\mathbf{k}
^{\prime}$, the initial and final states are not eigenstates of the spin
projection operator $S_{z}$ so the process relaxes the spin. The rate of
relaxation due to the EY process is well known to be of the form: $1/\tau
_{EY}=\alpha_{EY}T^{2}/\tau_{p}(T)$ where $\alpha_{EY}$ is a material
dependent parameter and $\tau_{p}$ is the momentum relaxation
time.\cite{chazalviel} We estimate $\alpha_{EY}(th)=4.6\times10^{-15}$
K$^{-2}$. \ The Bir-Aronov-Pikus (BAP) mechanism\cite{bir} arises from the
scattering of electron and holes. This relaxation mechanism is commonly
considered to be negligible in $n$-type materials like those under
consideration here since the number of holes is small.\cite{song} \ The
D'yakonov-Perel' (DP) mechanism \cite{dyakonov} arises from the ordinary
scattering of conduction-band states. \ Since this has not previously been
calculated in a wurtzite structure, we devote the next section to it. \ This
calculation yields an expression for $\tau_{c}$ as a function of temperature.

$\tau_{ls}$ and $\tau_{ld}$ are due to non-spin-conserving anisotropic
exchange (Dzyaloshinski-Moriya) interactions.\cite{dzyaloshinskii,moriya} The
anisotropic exchange term is important. It arises from spin-orbit coupling and
produces a term proportional to $\mathbf{d} \cdot\mathbf{s_{1}} \times
\mathbf{s_{2}}$ where $\mathbf{d}$ is related to the interspin separation and
the exchange integral between the wavefunction on sites 1 and 2. \ However, it
is not possible to calculate it in detail when the nature of the impurities is
not well known. \ We estimate the rate as $1/\tau_{DM}=\alpha_{DM}
(n_{imp,s}+n_{imp,d})$ where $n_{imp,s}$ and $n_{imp,d}$ are the total
impurity concentrations of the shallow and deep impurity respectively and
$\alpha_{DM}$ has a weak temperature dependence that we neglect. \ The main
contribution comes from the the overlap of the shallow impurity wavefunctions,
which we take to be hydrogenic, with the deep impurity wavefunctions, which we
take to be well-localized on an atomic scale. \ The details of how to estimate
the resulting relaxation may be found in Refs.
(\onlinecite{putikka,gorkov,kavokin}). The numerical value we find from theory
is $\alpha_{DM}(th)=1.12\times10^{-20}$ cm$^{3}$ ns$^{-1}$. When nuclei
possess nonzero magnetic moments, the hyperfine interaction between electron
and nuclear spin is a source of spin relaxation for localized
electrons.\cite{zutic} However, zero nuclear spin isotopes of Zn and O are
96\% and 99.5\% naturally abundant respectively. Therefore we rule out the
hyperfine interaction from being an observed relaxation mechanism in Ref. (\onlinecite{ghosh}).

\section{DP mechanism in wurtzite crystal structures}

The conduction band states undergo ordinary impurity and phonon scattering.
Each scattering event give a change in the wavevector $\mathbf{k}$, which in turn
changes the effective magnetic field on the spin that comes from spin-orbit
coupling. \ This fluctuating field relaxes the spin. \ The effective field
strength is proportional to the conduction band spin splitting. \ \ Bulk
zincblende crystals have conduction band splittings cubic-in-$k$ due to bulk
inversion asymmetry ( Dresselhaus effect ).\cite{dresselhaus} \ In addition to
cubic terms, bulk wurtzite conduction bands also possess spin splittings
proportional to linear terms in $k$ due to the hexagonal $c$ axis which gives
bulk wurtzite a reflection asymmetry similar to the Rashba
effect.\cite{rashba, voon, fu, weber} \ We can write the spin-orbit
Hamiltonian to include both the Rashba and Dresselhaus terms:
\begin{equation}
H_{so}(\mathbf{k})=[\alpha\mathbf{\kappa_{1}}+\gamma\mathbf{\kappa_{3}}
]\cdot\mathbf{\sigma} \label{eq:hamiltonian}%
\end{equation}
where $\kappa_{1}=(k_{y},-k_{x},0)$ is linear-in-$k$, $\kappa_{3}=(k_{||}
^{2}-bk_{z}^{2} )(k_{y},-k_{x},0)$ is cubic-in-$k$, $\mathbf{\sigma}
=(\sigma_{x},\sigma_{y},\sigma_{z})$ are the Pauli spin matrices, and $\alpha
$, $\gamma$ are spin splitting coefficients.\cite{voon, wang, fu} \ The
parameter $b$ is roughly equal to four for all wurtzite materials.\cite{wang}
\ Note that there is no spin splitting along the hexagonal axis ($z$).

\ The linear-in-$k$ term dominates and we can determine the spin relaxation
rate by following the treatment given by Pikus and Titkov in Ref.
(\onlinecite{oo}) which yields the following relaxation rates:
\begin{equation}
\frac{1}{\tau_{DP,ii}^{(1)}}=\tilde{\tau}_{p}\frac{4\alpha^{2}}{\hbar^{2}%
}(\overline{\kappa_{1}}^{2}-\overline{\kappa_{1,i}}^{2})\label{eq:relaxation}%
\end{equation}%
\begin{equation}
\frac{1}{\tau_{DP,i\neq j}^{(1)}}=\tilde{\tau}_{p}\frac{4\alpha^{2}}{\hbar
^{2}}\overline{\kappa_{1,i}\kappa_{1,j}}%
\end{equation}
where the overbar denotes angular averaging, and $i,j$ denote the Cartesian
components of $\mathbf{\kappa}_{1}$. The momentum relaxation rate is defined
as
\begin{equation}
\frac{1}{\tilde{\tau}_{p}}=\int_{-1}^{1}\sigma(\theta)(1-\cos\theta
)d\cos\theta
\end{equation}
where $\sigma(\theta)$ is the scattering cross section and $\theta$ is the
angle between initial and final $\mathbf{k}$.\cite{oo} \ In bulk wurtzite
$\overline{\kappa_{1,||}}^{2}=\overline{\kappa_{1,x}}^{2}=\overline
{\kappa_{1,y}}^{2}=k^{2}/3$, $\overline{\kappa_{1,z}}^{2}=0$, and in the
unstrained crystal, $\overline{\kappa_{1,i}\kappa_{1,j}}=0$ for $i\neq j$.
\ From Eq. (\ref{eq:relaxation}), we can write
\begin{equation}
\frac{1}{\tau_{DP,||}^{(1)}}=\frac{2}{\tau_{DP,z}^{(1)}}=\frac{4}{3}%
\frac{\alpha^{2}}{\hbar^{2}}\tilde{\tau}_{p}k^{2}=\frac{8}{3}\frac{m^{\ast
}\alpha^{2}}{\hbar^{4}}\tilde{\tau}_{p}E_{\mathbf{k}}%
\end{equation}
where $m^{\ast}$ is the electron effective mass and $E_{\mathbf{k}}$ is the
energy $\hbar^{2}k^{2}/2m^{\ast}$.\ This result can be Boltzmann averaged
(denoted by angle brackets) to obtain
\begin{equation}
\frac{1}{\tau_{DP}^{(1)}(T)}=\Bigg\langle\frac{1}{\tau_{DP,||}^{(1)}%
}\Bigg\rangle=\frac{8}{3}\frac{m^{\ast}\alpha^{2}}{\hbar^{4}}\langle
\tilde{\tau}_{p}E_{\mathbf{k}}\rangle=\alpha_{DP}^{(1)}\tau_{p}(T)T
\end{equation}
where $\alpha_{DP}^{(1)}=4m^{\ast}\alpha^{2}k_{B}/\hbar^{4}$ and $\tau
_{p}(T)=\langle\tilde{\tau_{p}}E_{\mathbf{k}}\rangle/\langle E_{\mathbf{k}
}\rangle=2\langle\tilde{\tau_{p}}E_{\mathbf{k}}\rangle/3k_{B}T$. The
temperature dependent momentum relaxation time, $\tau_{p}(T)$, can be
determined from electron mobility ($\mu_{e}$) measurements from $\mu_{e}
=e\tau_{p}(T)/m^{\ast}$ where $e$ is the charge of an electron. \ $\alpha$ has
been calculated \cite{voon} to be $1.1\times10^{-4}$ eV-nm which gives a
theoretical value of $\alpha_{DP}^{(1)}(th)=34.6$ K$^{-1}$ ns$^{-2}$.

\ Similarly, the cubic-in-$k$ term can be calculated to be
\begin{equation}
\frac{1}{\tau_{DP}^{(3)}(T)}=\frac{1}{\tau_{DP,||}^{(3)}(T)}=\frac{2}%
{\tau_{DP,z}^{(3)}(T)}=\alpha_{DP}^{(3)}\tau_{p}(T)T^{3}%
\end{equation}
where $\alpha_{DP}^{(3)}=80Q\gamma^{2}m^{\ast3}k_{B}^{3}/3\hbar^{8}$ where the
dimensionless quantity $Q$ depends on the type of scattering and is of order
unity. \ $\gamma$ has been calculated\cite{voon} to be $3.3\times10^{-4}$
eV-nm$^{3}$ which yields $\alpha_{DP}^{(3)}(th)=2.0\times10^{-4}$ K$^{-3}$
ns$^{-2}$.

The sample from which the momentum relaxation times $\tau_{p}(T)$ were
extracted \cite{look} was hydrothermally grown by the same company as Ghosh et
al. sample in Ref. (\onlinecite{ghosh}).

\section{Results and Discussion}

In Fig.1 we show that temperature dependence of $\tau_{s}$ as measured in a
bulk ZnO sample and our fit (using Eq. (\ref{eq:eigenvalue})) to the data.
\ It is seen immediately that the temperature dependence is not monotonic and
that this is well-reproduced by the theory. \ The reason is simple. \ At low
temperatures $T\ll|\varepsilon_{ls}|/k_{B}$ nearly all the electrons are in
localized states. \ These states relax by the temperature-independent
DM\ mechanism: $1/\tau_{ls}=1/\tau_{DM}$. \ This mechanism alone determines
the $T=0$ values. \ When $T$ approaches $|\varepsilon_{ls}|/k_{B},$ the deep
impurities are all occupied but the rest of the population is shared by
shallow localized and conduction band states. Initially, the conduction band
electrons have a longer spin lifetime because impurity scattering is frequent
at low temperatures so the DP mechanism that relaxes them is not very
effective. However, the DP mechanism increases rapidly as $T$ increases and
the $\tau_{s}$ curve turns around. \ At $T\gg|\varepsilon_{ls}|/k_{B},$ the
shallow impurity level is empty and the relaxation is dominated by the DP
mechanism in the conduction band: $1/\tau_{c}=1/\tau_{DP}^{(1)}(T)$.

At this point it is necessary to point out why only the linear-in-$T$ DP
mechanism is needed to explain the observed conduction spin relaxation. The
other two viable candidates (cubic DP and EY) for relaxation are much too weak
to explain the observed relaxation times in ZnO. We use the calculated values
for $\alpha_{DP}^{(1)}(th)$ and $\alpha_{DP}^{(3)}(th)$ in the previous
section to obtain the relative relaxation efficiencies between the linear and
cubic DP mechanism terms:
\begin{equation}
\frac{1/\tau_{DP}^{(1)}}{1/\tau_{DP}^{(3)}}=\frac{\alpha_{DP}^{(1)}%
(th)}{\alpha_{DP}^{(3)}(th)T^{2}}=\frac{1.73\times10^{5}~\text{K}^{2}}{T^{2}}%
\end{equation}
which demonstrates that the efficiency of the cubic-in-$T$ term does not
become comparable to the linear-in-$T$ term at temperatures below $416$ K
which is far above the temperature range investigated here. In fact the
cubic-in-$T$ term does not even reach one-tenth the efficiency of the
linear-in-$T$ term in the temperature range investigated here. For this reason
we can confidently ignore the cubic-in-$T$ DP mechanism term in our fit. The
crystal structure of ZnO therefore makes its spin relaxation qualitatively
different from spin relaxation in bulk $n$-GaAs. We also compare the
efficiencies of the DP and EY mechanisms:
\begin{equation}
\frac{1/\tau_{DP}^{(1)}}{1/\tau_{EY}}=\frac{\alpha_{DP}^{(1)}(th)\tau_{p}%
^{2}(T)}{\alpha_{EY}(th)T}=\frac{7.5\times10^{15}\tau_{p}^{2}(T)~\text{K
ns}^{-2}}{T}.
\end{equation}
Even if the momentum relaxation time taken to be unrealistically low, say $1$
fs, the DP mechanism is still nearly two orders of magnitude more efficient at
relaxing spins than the EY mechanism in the temperature range studied here.
Due to the drastic qualitative and quantitative differences between relaxation
mechanisms, we have unequivocally determined the relevant conduction electron
spin relaxation mechanism in ZnO.

\begin{figure}[ptbh]
\begin{centering}
	\includegraphics[scale = 0.3275,trim = 75 -20 35 50, angle = -90,clip]{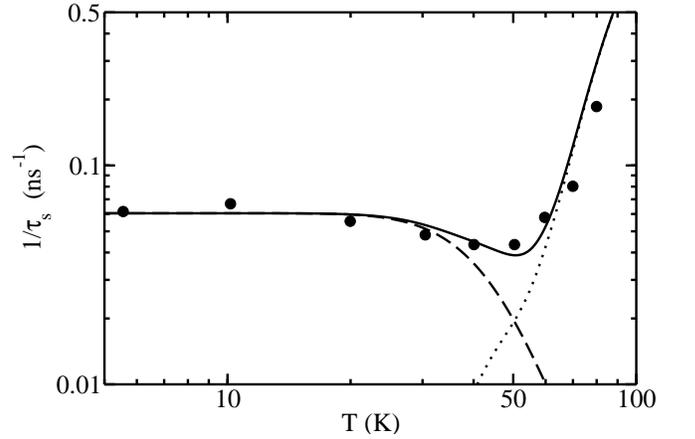}\label{fig:plot}
	\caption[]{Plot of $1/\tau_s$ vs. temperature. Points are experiment of Ref. \onlinecite{ghosh}. Dashed curve: $[n_{ls}/(n_c + n_{ls})] (1/\tau_{DM})$.
	 Dotted curve: $[n_{c}/(n_c + n_{ls})] (1/\tau_{DP})$. Solid curve: total $1/\tau_s$. $n_{imp,s} = 6.0 \times 10^{14}$ cm$^{-3}$,
	 $n_{imp,d} = 5.0 \times 10^{17}$ cm$^{-3}$, $\varepsilon_{ls} = -23$ meV, and $\varepsilon_{ld} = -360$ meV.}
	\end{centering}
\end{figure}

The fit of theory to the experimental data is clearly very good. \ We found
that no reasonable fit was possible using only a single impurity level, though
this worked very well for GaAs \cite{putikka}, so we used two levels. A good
fit by this method was possible by adjusting the coefficients $\alpha
_{DP}^{(1)}(exp)$ and $\alpha_{DM}(exp),$ and the binding energies
$\varepsilon_{ls},\varepsilon_{ld}$ and concentrations $n_{imp,s},n_{imp,d}$
of the two donors, subject to the constraint that the room temperature carrier
density should equal the measured\cite{ghosh} value of $1.26\times10^{15}$
cm$^{-3}$. \ Qualitatively, one finds that $n_{imp,d}\gg n_{imp,s}$ and
$|\varepsilon_{ld}|\gg|\varepsilon_{ls}|$ to get the right order of magnitude
of the relaxation at low $T.$ \ Physically, the deep impurity spins are
important because they relax the shallow impurity spins by the DM mechanism,
and the strength of the low $T$ relaxation implies that the deep impurities
must be quite numerous. \ Quantitatively, a least squares fit to the data of
Ref. (\onlinecite{ghosh}) yields $\alpha_{DP}^{(1)}(exp)=134.5$ K$^{-1}$
ns$^{-2}$, $\alpha_{DM}(exp) n_{imp,d} = 0.06$ ns$^{-1}$, \ $|\varepsilon
_{ld}|=360$ meV, $|\varepsilon_{ls}|=23$ meV, and $n_{imp,s}=6.0\times10^{14}$
cm$^{-3}.$

$\alpha_{DP}^{(1)}(exp)$ is about four times larger than the theoretical value
of $\alpha_{DP}^{(1)}(th)$ given above, possibly due to strain effects. \ We
also note that the values of $\tau_{p}$ that we used were taken from a
different sample.

If we take $n_{imp,d}$ to be near the highest values measured for the deep
donor (see below) then  $\alpha_{DM}(exp) = 12 \times10^{-20}$ cm$^{3}
$ns$^{-1}$ is about one order of magnitude larger than the theoretical
estimate $\alpha_{DM}(th)$ given above. In view of the very poor understanding
of the impurity wavefunctions, and the exponential dependence of $\alpha_{DM}$
on the overlaps, this is perhaps not too disturbing. 

The presence of a shallow donor and a very deep donor has been seen in
hydrothermally grown ZnO samples of the type investigated here.\cite{grossner,
seager} Donor concentrations up to nearly $5.0\times10^{17}$ cm$^{-3}$ (
$n_{imp,d}$) have been measured for donors $330-360$ meV ($\left\vert
\varepsilon_{ld}\right\vert )$ deep.\cite{kassier, grossner, seager} Donors as
shallow as $13-51$ meV ($\left\vert \varepsilon_{ls}\right\vert $) have been
measured\cite{wenckstern} at lower concentrations $\sim5.0\times10^{14}$
cm$^{-3}$ ($n_{imp,s})$. Comparison with our values indicates that the
parameters extracted from the fit are very reasonable for this material.

From this analysis, we predict that in ZnO samples with fewer deep impurities,
the relaxation time at low temperatures can be increased. As the impurities of
ZnO vary greatly between different growth techniques\cite{look2}, this
prediction could be tested by further optical orientation experiments on
different samples.

\section{Conclusions}

We have found that $\tau_{s}$ in bulk ZnO can be understood by invoking
previously known spin relaxation mechanisms. \ The dominant mechanisms in the
material turn out to be the DP (scattering) relaxation of the conduction
electron spins for $T>50$ K and the DM (anisotropic exchange) mechanism for
the localized spins for $T<50$ K. \ In addition, it is very important to
include the cross-relaxation between localized and conduction states
previously proposed for GaAs. \ These physical ingredients explain
quantitatively the relatively fast relaxation at low temperatures as being due
mainly to the DM mechanism which in turn depends on having both deep and
shallow impurity states. \ At high temperatures, the conduction states are
dominant, and the DP mechanism gives an excellent fit to the data. \ The
combination explains the very surprising non-monotonic temperature dependence
of $\tau_{s}$.

Finally, there are two aspects of the data in Ref. (\onlinecite{ghosh}) that
we have not addressed here: the applied magnetic field dependences on the spin
relaxation and the spin relaxation observed in ZnO epilayers. We plan on
addressing the former issue in a future publication. As for the latter issue,
the epilayers are doped three to four orders of magnitude higher than in the
bulk case. At such high dopings, spin glass effects become important and
localized donor states coalesce to produce donor bands; we do not expect our
theory to be applicable in such a regime.

The theory has now been sufficiently developed that optical orientation
experiments can actually serve as a characterization tool for doped
semiconductors, giving information about the binding energies and
concentrations of the electrically active impurities in $n$-type materials.

We would like to acknowledge useful discussions with S. Ghosh. \ Financial
support was provided by the National Science Foundation, Grant Nos.
NSF-ECS-0523918 (NH and WP) and NSF-ECS-0524253 (RJ).\ \ 

\ \ 

\ \ \ 

\
%\bibliography{refs}

\

\end{document}